\documentclass[aps,prb,showpacs,floatfix,amsmath,amssymb,
reprint,superscriptaddress]{revtex4-1}
\usepackage{color}
\usepackage{graphicx}
\usepackage{natbib}
\usepackage{epsfig}
\usepackage{setspace}
\usepackage{amsmath,amssymb}
\usepackage{verbatim}
\usepackage{bibentry}
\usepackage[utf8]{inputenc} 
\usepackage[T1]{fontenc}

\def\nnu{{\nonumber}}

\begin{document}
\title{Fragility of the Kondo insulating gap against disorder: relevance to recent puzzles in topological Kondo insulators}


\author{Sudeshna Sen}
\affiliation{Tsung-Dao Lee Institute, School of Physics and Astronomy, Shanghai Jiao Tong University, Shanghai 200240, China}
\affiliation{University College Dublin, Befield, Dublin-4, Ireland}
\altaffiliation{corresponding email: sudeshna.sen@ucd.ie}
\author{N. S. Vidhyadhiraja} 
\affiliation{Theoretical Sciences Unit, Jawaharlal Nehru Centre for Advanced 
Scientific Research, Bangalore-560064, India}
\author{Eduardo Miranda}
\affiliation{"Gleb Wataghin" Institute of Physics, University of Campinas - Unicamp, 13083-859, Campinas-SP, Brazil}
\author{Vladimir Dobrosavljevi\'c}
\affiliation{Department of Physics and National High Magnetic Field Laboratory,
Florida State University, Tallahassee, Florida 32306, USA}
\author{Wei Ku}
\altaffiliation{corresponding email: weiku@mailaps.org}
\affiliation{Tsung-Dao Lee Institute, School of Physics and Astronomy, Shanghai Jiao Tong University, Shanghai 200240, China}
\affiliation{Key Laboratory of Artificial Structures and Quantum Control (Ministry of Education), Shanghai 200240, China}

\begin{abstract}
Kondo insulators are strongly correlated systems in which a clean insulating gap emerges only at very low temperature due to many-body effects invovling localized $f$-electrons.  However, certain Kondo insulators, like SmB$_6$ and Ce$_3$Bi${_4}$Pt${_3}$, display metallic behaviors at extremely low temperature, that have defied current understanding.  Recent advances in topological effects in materials has raised the attention on the protected surface states in these ``topological Kondo insulators'' as a potential resolution to some of the puzzling behaviors.  Here we resolve these puzzles via a different route, by showing that the emergent Kondo insulating scale is extremely vulnerable against moderate degree of disorder, such that the gap is filled with small number of states.  Therefore, the real samples are probably never truly insulating and this in turn compromises the essential building block of topological considerations.  Our results suggests strongly that systems like the Slater insulators would be a more promising direction to extend the realm of topology to strongly correlated systems.
\end{abstract}
\maketitle
\section{Introduction}
In recent years, topological Kondo insulators~\cite{Coleman_review} have emerged as a new class of materials where both the physics of strong electron correlations and/or topology could play a significant role. 
Interestingly, however, in great contrast to typical topological insulators, these systems also exhibit seemingly contradictory behaviors, for example, a low-temperature metallic specific heat \cite{sp_ht_old,PhysRevX.4.031012,Fisk_sp_ht} and an insulating-like activated transport at high temperatures~\cite{Allen1979,Coleman_review,foreword,Gorshunov}.
Even more puzzling is the observation of a saturated low-temperature resistivity \cite{Allen1979,sp_ht_old, Coleman_review,foreword,Gorshunov}.
Naturally this can be interpreted from the conducting topological surface states~\cite{tuning_bulk_and_surface_conduction,kim2013surface,PhysRevB.91.085107,Park14062016,2014NatCoXu,2013NatCoZHasan,2014Fisk,lee2019perfect}. However, the origin of the experimental observations are also unclear, demonstrating extensive sample dependence or influence from the specific experimental design, indicating the possible role of extrinsic effects~\cite{PhysRevB.100.035435,PhysRevLett.122.166401,das2019comment}.
Moreover, very recent observations strongly suggest bulk conduction instead~\cite{eo2019transport,Lieaap8294}, and advocate the essential role of disorder~\cite{Bulkacconduction2016,eo2019transport,GABANI201517,field_dependent_Mcqueen}.

Thereby, several experimental observations in some of these systems have called for unconventional theoretical interpretations.
Not only are the low temperature thermodynamic properties of this bulk insulator at odds with conventional knowledge, this material was recently found to exhibit 3D quantum oscillations~\cite{Tan287,2018arXiv180303553V} typically associated with metals.
To date, while several propositions have been attempted to understand such puzzling observations, including charge neutral quasi-particles~\cite{Cooper,Senthil2017,Senthil2018,2015arXiv150703477B} and conducting surface states~\cite{yu2015model,kim2014topological,foreword,PhysRevX.4.031012}, no overall consistent picture has been obtained.

One long-standing key issue is the role of disorder~\cite{miranda1996kondo}.
Even before realizing the connection with topological characteristics, disorder had already been speculated to be responsible for the saturation of low-temperature resistivity~\cite{Allen1979}.
More recent studies further found significant sample variation in the low-temperature properties of these systems, depending on the synthesis methods and seed materials used \cite{valentine2016breakdown,eo2019transport,screened_moments,field_dependent_Mcqueen}. 
Careful characterization of such samples indeed indicates that even in the nominally purest samples, minimal yet detectable amount ($< 1$\%) of disorder~\cite{valentine2016breakdown} is present, the amount of which appears to be correlated with their saturated value of resistivity~\cite{valentine2016breakdown} and specific heat~\cite{sp_ht_old}. 
But, how can such a small amount of disorder overcome the insulating gap of 0.01eV scale, and are they necessarily itinerant? It is exactly this question that we address in this work and provide a microscopic physical mechanism explaining this issue.

\section{Model}
To illustrate this generic characteristic of all Kondo insulators, we use a minimal model known as the periodic Anderson model (PAM), consisting of itinerant non-interacting $d$ orbitals with on-site energy $\epsilon_{dj}$ and highly localised interacting $f$ orbitals with on-site Coulomb repulsion $U$ and on-site energy $\epsilon_{f}$.
Additionally, the $d$ and $f$ orbitals hybridize via a local coupling $V$, such that, the Hamiltonian is given in standard notation by, $H=H_d+H_f+H_{hyb}$, where 
\begin{align}
  H&=-\sum_{\langle ij\rangle\sigma}\lbrack t_{ij} d_{i\sigma}^\dagger d_{j\sigma}
       + H.c.\rbrack + \sum_{j\sigma} (\epsilon_{dj}-\mu)d_{j\sigma}^\dagger d_{j\sigma}\nonumber\\ 
  &+ \sum_{j\sigma} \epsilon_{f}f_{j\sigma}^\dagger f_{j\sigma} 
       + U \sum_j n_{fj\uparrow}n_{fj\downarrow}
  +V \sum_{j\sigma}\lbrack d_{j\sigma}^\dagger f_{j\sigma} + H.c.\rbrack,
 \label{eq:PAM}
\end{align}
where $\langle ij\rangle$ denotes nearest-neighbour hopping and $t_{ij}=t/\sqrt{N}$ (in the limit when lattice coordination number, $N\to\infty$).
The last term in Eq.~\eqref{eq:PAM}, $H_{hyb}$ results in the formation of a hybridization gap, $\Delta_g$, in a $1/2$-filled lattice of the above model. The $1/2$-filling condition in a clean Kondo insulator is respected when the $f$-orbital occupancy, $n_f$ and the $d$-orbital occupancy, $n_d$ together sum to, $n_f+n_d=2$ and the chemical potential $\mu$ lies inside the gap. 
In a disordered system, these occupancies are replaced by their disorder averaged values, such that for a disordered Kondo insulator, the former relation should read as, $\langle n_f\rangle+\langle n_d\rangle=2$, where $\langle ... \rangle$ denotes disorder averaged values. 

In the following sections we consider the effects of finite disorder in a symmetric Kondo insulator. Before delving into the fully interacting model, we first illustrate the role of disorder in the $U=0$ limit of the above model. 

\section{Results}
\begin{figure}[htp!]
  \includegraphics[clip=,width=0.8\columnwidth]{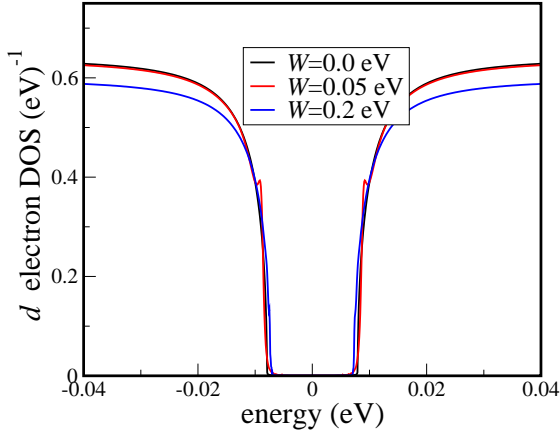}
  \caption{{\bf Effect of disorder in a non-interacting band insulator:} The ground state density of states (DOS) of a non-interacting band insulator, is plotted as a function of energy, for different values of standard Anderson-type disorder in site-energies. $W$ is a measure of the fluctuation of site energy with values $W=0,\,0.05,\,0.2$eV. Clearly, disorder has negligible effect on the hybridization gap straddling around the chemical potential located at zero energy. The other parameters used are $U=0.0$eV and $V=0.089$eV.}
  \label{fig:U0}
\end{figure}

{\it Disorder in a band insulator}- Figure~\ref{fig:U0} gives an example of typical effect of minute amount of disorder on band insulators, containing a bulk insulating gap of full width $\Delta_g\approx 0.016$eV, obtained by putting $U=0$ and $V=0.089$eV in Eq.~\eqref{eq:PAM}. We simulate the disorder effects via standard Anderson-type fluctuation of $d$-site energies randomly distributed according as a Gaussian probability distribution function, $P(\epsilon_{di})=\frac{1}{\sqrt{2\pi W^2}}\mathrm{exp}\lbrace-\frac{1}{2}\epsilon_{di}^2/W^2\rbrace$, with variance or disorder strength, $W$ and mean zero, such that $\langle n_d\rangle=1$ on average.
Furthermore, the band-filling constraint by fixing $n_f\approx1$ uniformly on all sites, while varying $n_d$ locally via a spatially varying parameter $\epsilon_{di}$, drawn from the distribution $P(\epsilon_{di})$.
Thus, within this model, $W$ would not only signify the local fluctuations in the $d$-levels but  additionally imply fluctuations in the local density of states (LDOS) of the conduction electrons~\cite{tanaskovic2004effective} which would otherwise be missed in the kind of mean field theory adopted in this work.
In all our subsequent discussions based on the electronic density of states, we choose the $d$-electrons, as they are the ones involved in the transport.

Before moving to the results, we would comment on how robust could the results of our model calculations be, to the details of the microscopic model for disorder. Detailed investigations~\cite{tanaskovic2004effective}, have convincingly revealed that the resulting power-law form of the distribution of Kondo temperatures is remarkably robust and insensitive to any specific features of the realistic disorder distribution. Nonlocal disorder correlations (``Friedel oscillations") lead to significant disorder renormalization in the conduction band, which assumes a generic Gaussian form, providing strong support for the usage of an effective model such as the one used here.

In Figure~\ref{fig:U0} we demonstrate the $d$-electron density of states for such site energy fluctuations up to $W=0.2$eV.
Expectedly, one finds only negligible effects near the gap edge in the resulting density of states.
That is, the gap is extremely robust against such weak disorder.
This means that even for an underestimated resistivity (by assuming all the states are itinerant), one would still obtain a large insulating-like resistivity at low temperature, without saturation, unlike that observed in Kondo insulators like SmB$_6$.
It turns out that the resolution of the above issues lies in a novel characteristic of Kondo insulators in general (topological or not), in contrast to typical band insulators, namely an extreme sensitivity of local gap features against disorder.

{\it Disorder in a Kondo insulator}- The model represented by $H$ is known to capture the non-trivial local physics of Kondo screening at very low temperature in the regime when $Ut\gg V^2$, where the very low-energy physics is controlled by the emergence of bound singlet states.
These entangled spin-singlet states, also known as the Kondo singlets, are composed of anti-ferromagnetically coupled $d$ and $f$ electrons via an spin exchange coupling $J\sim V^2/U$ and a binding energy, so-called Kondo scale $\omega_K$.
The most dramatic characteristic of the Kondo screening is the exponential (many orders of magnitude) suppression of its energy scale from that of the Hamiltonian ($t$, $V$, and $U$), $\omega_K\propto e^{-1/(\rho_0J)}$, where $\rho_0$ represents the ``bare'' density of states of the $d$ electrons at the chemical potential.
Consequently, near half-filling (number of $d$ electrons $n_d$ and $f$ electron $n_f$ sums to 2 per atom), the coherent charge gap opening related to the periodic occurrence of the Kondo singlets in a lattice is also very small, of the order of the Kondo scale, $\Delta_g\sim \omega_K$.
A Kondo insulator is thus formed at exactly half-filling, when the chemical potential falls inside the charge gap.

Note that $H$ traditionally describes a Kondo insulator with strictly on-site hybridization between the $d$ and $f$-orbitals. In a realistic system, however, the spatial dependence of the hybridization term may not be ignored because such a structure could be responsible for generating the topological characteristics.  However, this consideration is nearly orthogonal to the physics of local Kondo physics. To illustrate this point, let’s consider the following representative effective hybridization (on a two dimensional lattice) expanded around the special -points with band inversion across the Kondo gap: $V(\mathbf{k})\sim\alpha+\beta(k_x+ik_y)\sim\alpha+\beta(\sin(k_x)+i\sin(k_y))$.
As long as $|\alpha|$ is smaller than $|\beta|$, such a hybridization would produce a non-trivial topological structure. To more rigorously capture the short-range hybridization, one can construct a properly symmetric superposition of $d$-orbitals from all sites surrounding the $f$-orbital site, to represent the same $d$-orbital degrees of freedom, in a manner similar to the well-known Zhang-Rice singlet, as demonstrated in ~\cite{Wei_2013}. This way, one can absorb the most essential effects to the local Kondo physics through a local hybridization between the $f$-orbitals and the new (larger) symmetrized $d$-orbitals.
This includes a proper symmetric phase structure that leads to the topological characteristics of SmB$_6$. In fact, the community has recently witnessed  a renewed focus in the designing of model topological Hamiltonians for SmB$_6$ (and systems alike)~\cite{Dzero2012,TKI_model,Vojta_TI_models2015,Vojta2014,2018arXiv180905850G,2018JPSJKawakami}. Note, however, that this issue of global topology is a lower-energy physics emerging below the higher-energy local Kondo gap scale. Therefore, as far as the topological Kondo insulators are concerned, the development of the bulk Kondo insulating gap should display nothing peculiar, as compared to conventional Kondo insulators. 
\begin{figure}[htp!]
  \includegraphics[clip=,scale=0.45]{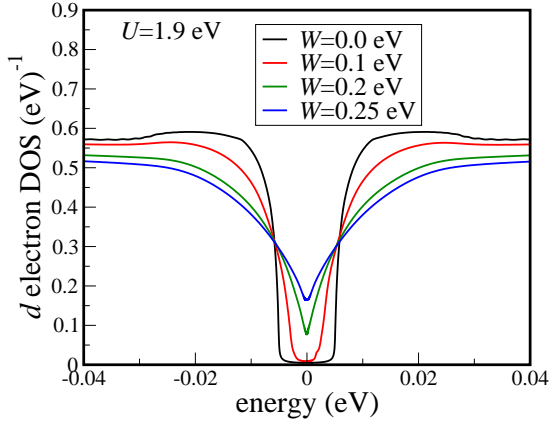}
  \caption{{\bf Effect of disorder on the Kondo insulator density of states:} The zero temperature density of states (DOS) of the $d$ electrons of a Kondo insulator is plotted as a function of energy, for different values of $d$ site energy fluctuation. The magnitude of fluctuation is represented by $W$. In contrast to the $U=0$ case (Fig.~\ref{fig:U0}), the Kondo insulating gap of a clean Kondo insulator ($W=0$) starts filling up at minimal disorder strengths ($W=0.1$) rapidly evolving into a 
  pseudogap $(W=0.2,\;0.25)$eV, with a non-zero density of states at the chemical potential. The parameters used 
  for this figure are $U=1.9$, $V=0.44$eV to generate a clean Kondo insulating gap of full width $\Delta_g\approx0.01$ eV.}
  \label{fig:U1.9}
\end{figure} 

Thus, for the purposes of demonstrating the genuinely, non-trivial role of Kondo disorder on the transport and thermodynamic properties of Kondo insulators, we consider a spatially local $V$. We solve the model described in Eq.~\eqref{eq:PAM} utilizing the dynamical mean field theory (DMFT) framework~\cite{DMFT} for tackling the many-body effects due to the strongly correlated $f$-electrons and the coherent potential approximation (CPA) to understand the effects due to disorder. The DMFT is formally exact in the limit of infinite dimensions, where the interaction self-energy becomes purely local in space but retains full temporal dynamics of the interaction self-energy. The CPA applied to treat the disorder is a mean-field approach to determine the {\it disorder-averaged} effective DMFT medium seen by the $f$-electrons, $\Delta_{ave}(\omega)$, obtained by solving a set of self-consistent equations as outlined in Appendix~\ref{app1}. The CPA is known to give reliable results for regimes where disorder induced localization effects are negligible~\cite{miranda1996kondo}. In order to reliably capture the exponentially suppressed Kondo energy scale, we obtain the local self-energy of the PAM using the local moment approach within the framework of DMFT~\cite{logan2002finitetempdynamics,Raja2_2005,vidhyadhiraja2003dynamics,vidhyadhiraja2004dynamics,logan2005dynamics}. For further details on the implementation of the algorithm within the disorder framework we redirect the reader to Appendix~\ref{app1}. Although, these results should hold true for any high dimensional lattice where the local self-energy serves as a reliable approximation, for all the results demonstrated here we have used the Bethe lattice density of states for the free conduction band, given by, $\rho_0(e)=\frac{2}{\pi t}[1-(e/t)^2]^{1/2}$, $2t$ is the conduction electron full bandwidth (and $t=1$ for all the calculations presented).

Now we will show that unlike the typical charge gap in band insulators, such a dramatically suppressed emerged gap (and the local Kondo scale) is extremely sensitive to disorder. It is worth mentioning that the general fragility of correlated systems to disorder was to some extent discussed in past work~\cite{miranda1996kondo,miranda1997disorder,miranda2001griffiths,tanaskovic2004effective}. However, the precise consequences for the thermal behavior of such disordered Kondo insulators was not addressed. Other work restricted to binary type (Kondo holes/ligand disorder) had also been reported previously~\cite{Anders,Raja_Pramod} including a recent work demonstrating the role of in-gap bound states due to non-magnetic impurties ~\cite{PhysRevB.101.094101}; the latter models however cannot produce a broad distribution of Kondo scales, a key entity for observing the `fragile' Kondo insulator scenario as discussed below.
This fragility of the Kondo insulating gap against disorder due to a broad distribution of Kondo scales provides a natural explanation of the above puzzles.
We simulate the weak disorder in real materials by adding to this model Anderson-type disorder in the $d$-electron site energies, with fluctuations $\sim 0.2$eV, similar to that used for the non-interacting case demonstrated in Fig.~\ref{fig:U0}.

Instead of delving into the detailed low energy dynamics that would dictate this physical situation, a straight look at the resulting conduction electron density of states (equivalently the disorder averaged $c$-electron spectral function) illustrated in Figure~\ref{fig:U1.9} immediately reveals the `fragility' of the Kondo gap in great contrast to its non-interacting counterpart (Figure~\ref{fig:U0}).
With the incorporation of disorder, the prominent charge gap rapidly fills out representing a `soft' gap at $W=0.1$eV and evolving into a `pseudogap' with finite spectral weight at the chemical potential, with a meagre increase of $W$ to $0.2$eV. 

\begin{figure}[htp!]
\includegraphics[width=\columnwidth]{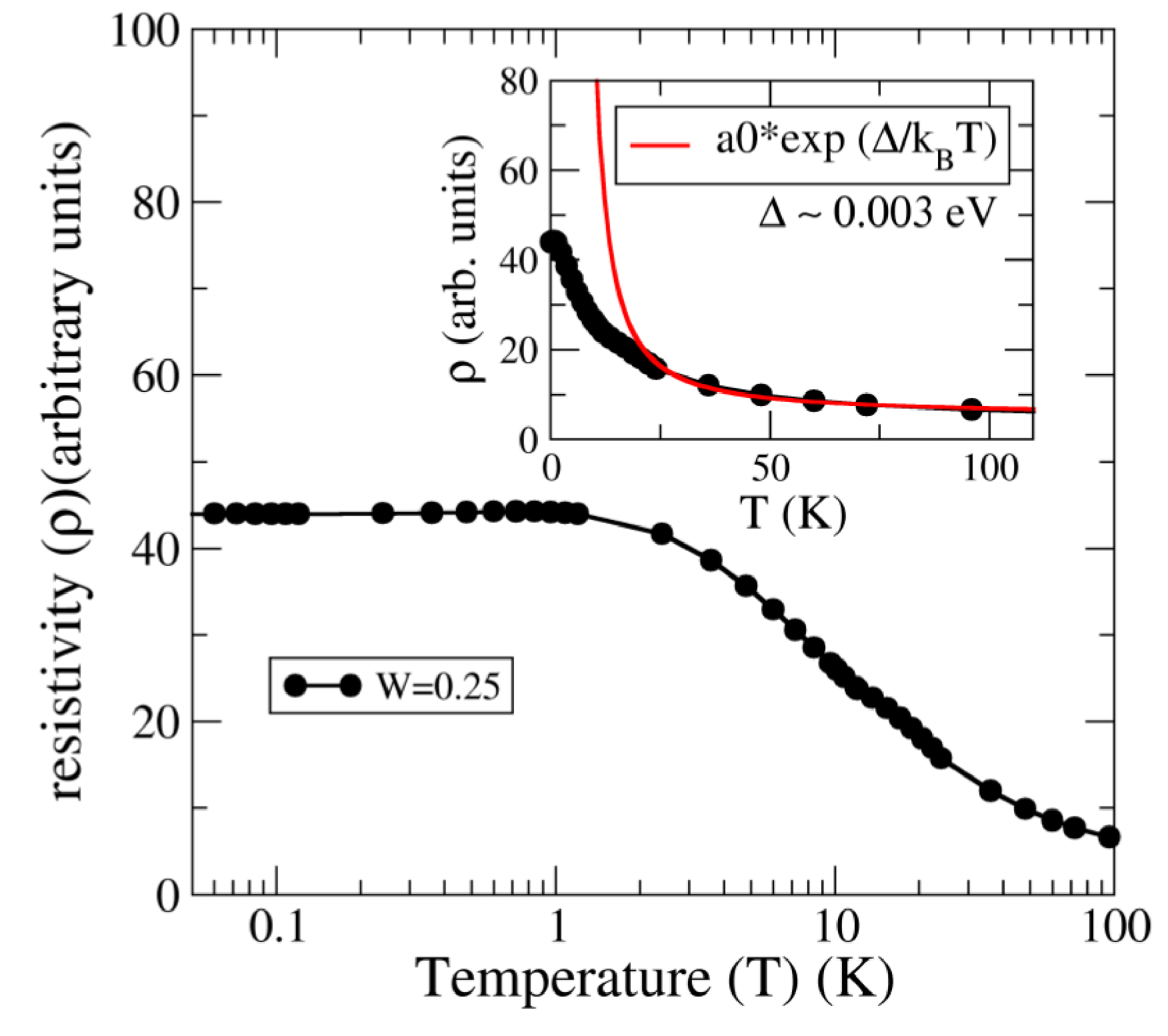}
  \caption{{\bf D.C. resistivity:} (Main) The temperature ($T$) dependent D.C. resistivity ($\rho(T)$) is plotted, for $W=0.25$eV, with $T$ being plotted on a log scale. In agreement with experimental observations, a resistivity plateau sets   in for temperatures, $T\lesssim 2$ K, crossing over to a regime of activated transport characteristic of Kondo insulators. (Inset) In the inset we plot $\rho(T)$ on a linear scale to highlight the activated transport for $T\gtrsim 20$K with a transport gap, $\Delta\sim0.003$ eV.}
  \label{fig:rho}
\end{figure}

{\it Transport and thermodynamics}-
The occurrence of such `pseudogap' feature in the ground-state, bulk spectral function would not only induce a plateau in low T, D.C. resistivity ($\rho(T)$) of such a system, but also result in a high temperature insulator like activated transport as we now demonstrate. 
In main panel of Fig.~\ref{fig:rho} we plot $\rho(T)$ on a linear-log scale for $W=0.25$. The saturation knee occurs at a $T\sim 2$K in close agreement to experimental reports on SmB$_6$. 
In the inset we plot $\rho(T)$ on a linear scale to highlight the activated transport regime at higher $T$'s, characteristic of insulators. 
Our model of a disordered Kondo insulator produces an activated transport for $T\gtrsim 20$ K with a transport gap, $\Delta\sim0.003$eV which is of the same order as that reported in experiments on SmB$_6$ ~\cite{FISK1995798,FISK1996409,FISK6842,Eo12638}.

Note that the activation behavior should correspond to the size of the pseudogap, exactly as in any clean insulator. It is not until the pseudogap-related thermal activation becomes weaker at low temperature,   due to the presence of the few disorder-induced in-gap states, that the resistivity starts to saturate. Since the clean Kondo gap is controlled by the local Kondo physics (local density of states and hybridization), its size and the associated activation behavior should be quite robust (until saturation). That is, the cleaner the sample is, the lower the temperature at which the resistivity saturation should  take place, and also the higher the magnitude of the saturated resistivity. This expectation is consistent with the experimental findings of Ref.~\cite{eo2019transport} where the most stoichiometric (i.e. the cleanest) samples demonstrated a rather robust clean Kondo gap.  

In the following discussion we now show how our model for disordered Kondo insulators not only gives rise to such (weakly) metallic behavior, with finite conductivity at $T=0$, but also a modified thermodynamic response, in consistence with the `puzzling' experimental observations in several Kondo insulators.

\begin{figure}[htp!]
\includegraphics[clip=,scale=0.5]{Figure4.eps}
  \caption{{\bf Electronic specific heat:} (Main) The specific heat ($C$) in units of the universal gas constant ($R$), is plotted as a function of temperature ($T$), for a fixed disorder strength, $W=0.25$eV. Clearly, the low $T$ specific heat has an appreciable linear component as also observed in experiments. We observe two regimes with a dominant linear coefficient $\gamma=0.1,\;0.05$K$^{-1}$ respectively, the change of slope being accompanied by a knee at $\sim 2$ K. The lowest temperature reported  in the available experimental data for the specific heat of SmB$_6$ is $\sim 2$ K. As demonstrated in the figure, our model predicts a $\gamma\approx 0.05 R\sim 100$ mJK$^{-2}$mol$^{-1}$ in the regime $2-8$ K. (Inset) The linear coefficient, $\gamma$, is plotted as a function of $T$ on a linear-log scale.}
  \label{fig:Cv}
\end{figure}

Among the several puzzling observations in KI's SmB$_6$ and Ce$_3$Pt$_3$Bi$_4$, is the large low $T$ specific heat contribution with a dominant $T$-linear contribution, shown to be predominantly a bulk property~\cite{sp_ht_old,Fisk_sp_ht,2018arXiv180303553V,hartstein2018fermi}. 
We thus turn our focus on to the thermodynamic response due to the electronic specific heat, $C(T)$, and demonstrate how our microscopic model reproduces such an experimental observation. In order to obtain $C(T)$ we first evaluated the total energy of the disordered system (within the framework of CPA-DMFT, see Appendix~\ref{appc} and Ref~\onlinecite{Costi_sp_ht} for more details) and then calculated its derivative w.r.t. temperature\footnote{The $T\to0$ part of the $C(T)$ data may incur spurious features due to the finite interpolation of the total energy on to a denser temperature grid. In this case the best method is to fit the total energy}.
Figure~\ref{fig:Cv} (main panel) shows the computed specific heat ($C(T)$) in units of the universal gas constant, as a function of temperature with $R\approx 8.3$ JK$^{-1}$mol$^{-1}$, for $U=1.9$eV and $V=0.44$eV. 
The specific heat clearly depicts a low $T$ linear behavior until $T\sim 2K$ crossing over to higher temperatures with a similar linear trend, albeit with a reduced linear coefficient. Thus, the observed $C(T)$ behaves as $C(T)=\gamma R T$ exhibiting an enhancement in $\gamma$ for $T\lesssim 2K$ as highlighted in figure~\ref{fig:Cv}(inset). 
It should be noted that the most recent $C(T)$ measurements exhibit a significant sample dependence with $\gamma\sim 10-50$ mJ K$^{-2}$mol$^{-1}$~\cite{PhysRevX.4.031012,Fisk_sp_ht,Gabani2002,hartstein2018fermi}. 
As shown, our calculations successfully bring out the general trends observed in the experiments concerning candidate TKI's, namely SmB$_6$ and Ce$_3$Pt$_3$Bi$_4$, in regards to the bulk thermodynamic response. 
The thermodynamic response is also in consistence with the observed transport in these systems. 
Nevertheless, the $\gamma$ value predicted by our calculations is $\sim 100$ mJ K$^{-2}$mol$^{-1}$ (for 2 K$<\;T<$ 10 K) is slightly overestimated than experimental reports.

\begin{figure}[htp!]
  \includegraphics[clip=,scale=0.5]{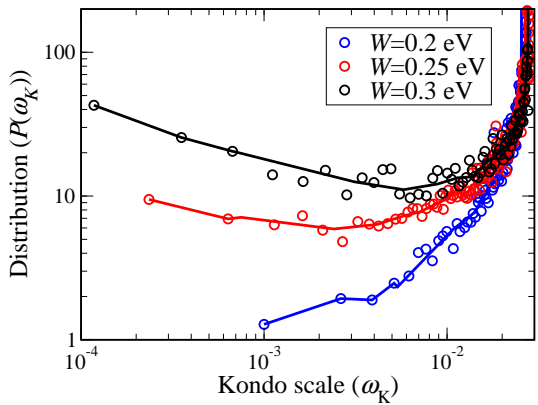}
  \caption{{\bf Distribution($P$) of Kondo scales($\omega_K$): $P(\omega_K)$} The distribution of local 
Kondo scales ($P(\omega_K)$) computed for different strengths of fluctuation in the Anderson-type quenched disorder,{\it viz}, $W=0.2,\,0.25,\,0.3$eV. The solid lines are a guide to the eye. The bare Hamiltonian parameters used are $U=1.9,\;V=0.44$eV.}
  \label{fig:P_TK}
\end{figure}
The microscopic mechanism behind the fragile nature of a Kondo insulator against disorder (and the related pseudogap formation) can be understood as the following.
The randomness introduced by site energy fluctuations induce a spatially non-uniform screening of the localised $f$-electrons.  
More precisely, the strength of this {\it local} screening depends on the local hybridization function or quantum-mechanical coupling ($\Delta(\epsilon_d)$) between the $f$-electrons and the $d$-electrons. 
The {\it local} Kondo scale ($\omega_{K}$) quantifying this process of Kondo screening is thereby given by $\omega_{K}\propto exp(-1/\Delta(\epsilon_d))$. 
The $f$-electrons thus become extremely sensitive to this microscopic fluctuation in the $d$-electron local density of states induced by the local disorder. 
Consequently, a broad distribution of `exponentially' small {\it local} Kondo scales emerge as illustrated in Figure~\ref{fig:P_TK}, where we demonstrate $P(\omega_K)$ for three such closely spaced disorder values, namely, $W=0.2,\,0.25,\,0.3$eV. Within the local moment approach impurity solver used in this work, the Kondo scale, $\omega_K$ is synonymous with the low energy spin-flip scale, identified as the position of the peak in the transverse spin polarization propagator that quantifies the transverse spin-flip processes responsible for Kondo effect. For more details the reader is referred to Appendix~\ref{app1}.
As shown in Figure~\ref{fig:P_TK}, these emerged scales (spanning from $10^{-4}$ to $10^{-2}$eV) are orders of magnitude smaller than the clean lattice coherence scale. 

Such a broad distribution of Kondo scales provides a broad avenue of extremely low energy scales over which  the underlying Kondo insulator density of states is dramatically influenced. Note that it is very important to precisely determine the ‘interaction renormalizations’ via the emergent, exponentially small, distribution of Kondo scales to demonstrate that even a minute amount of disorder could kill the Kondo insulating gap. Furthermore, our quantitative prediction prevails in the regime of relevance to topological Kondo insulators.
The Kondo gap that would straddle the Fermi level in the clean Kondo insulator would now move above or below the Fermi level depending on the $\lbrace\epsilon_{di}\rbrace$, and the amount of spectral weight acquired around the Fermi level, would depend on the underlying local Kondo scale which could be exponentially small as seen from Figure~\ref{fig:P_TK}.
In fact, the local depletion or accumulation of local $d$-orbital charge, $n_d$, may also reflect as many body quasi-particle or low energy resonances in the $f$ and $c$ spectral functions, in the vicinity of the Fermi level (for details see  Figure~\ref{fig:DOS_ed} in Appendix~\ref{app1}) on energy scales of the order of the underlying Kondo scale. 
Thus in great contrast to a non-interacting band insulator the Kondo insulating charge gap is extremely fragile in the presence of such quenched disorder, producing dramatically large effects on the low energy density of states with nominal site energy fluctuations.

It should be noted that for numerical reasons we cut off the tails of the bare Gaussian distribution, $P(\epsilon_d)$ such that we have a {\it bounded} distribution in effect. Thus the plotted $P(\omega_K)$ in Figure~\ref{fig:P_TK} depicts a range of $\omega_K$'s with a minimum $\omega_K$ and Fermi liquid behavior is expected to occur below this $\omega_K$. For example, for $W=0.25$ we expect Fermi liquid behavior for $\omega_K\lesssim 10^{-4}$eV (1K) in agreement with the specific heat depicted in Fig.~\ref{fig:Cv}. In experiments, it is difficult to quantify the amount and effects of disorder in a controlled manner and {\it apriori} unclear what the bare disorder distribution is. We believe that this cut-off of the {\it rare realizations} is justified since the amount of disorder required to close the gap is very small in comparison to the other energy scales of the bare model. A detailed correlation between the obtained $C(T)$ and the $P(\omega_K)$ and its evolution with increasing disorder is left as a future work.

\section{Discussion and Conclusions}
In summary, we demonstrate the extreme fragility of Kondo insulating gap against even modest amount of disorder, in dramatic contrast to conventional band insulators. Consequently, Kondo insulators generically develop a pseudo-gap that naturally accounts for the seemingly contradictory 
observations (metallic vs. insulating characteristics) in several topological Kondo insulators. 

In essence, our discovery of disorder-induced mid-(bulk)gap states and the sensitivity of local Kondo physics against disorder implies that even if there exist spin-locking (topological) surfaces states, they are no longer strictly protected against impurity scattering or hybridization with low-energy bulk states, unless the samples are devoid of any impurities or defects. Once the bulk Kondo gap, the building block of any topological insulator is undermined, any non-trivial physical role of the surface states will be seriously compromised, making them less relevant to the observable properties of the materials.

In a very recent work ~\cite{Qosc_in_gap_states}, the question of quantum oscillations from disorder-induced in-gap states in band insulators was studied within a phenomenological approach again showing strong evidence of the observation of quantum oscillations via disorder-induced in-gap states in low gap insulators, but this issue remains yet to be addressed for (topological) Kondo insulators.

\begin{acknowledgements}
SS and WK acknowledges the support from National Natural Science Foundation of China \#11674220 and 11447601, and Ministry of Science and Technology \#2016YFA0300500 and 2016YFA0300501. SS acknowledges discussions with Andrew Mitchell and research funding from the Irish Research Council Laureate Awards 2017/2018 through grant IR-CLA/2017/169. Work in Florida was supported by the NSF Grant No. DMR-1822258, and the National High Magnetic Field Laboratory through the NSF Cooperative Agreement No. DMR- 1157490 and the State of Florida. EM acknowledges a grant from CNPq (Grant N. 307041/2017-4). NSV acknowledges a RAK-CAM senior fellowship as well as funding from an SERB grant (EMR/2017/005398) and JNCASR.
\end{acknowledgements}
\appendix
\section{Calculation of the spectral function}
\label{app1}
In the main text we reported the analysis of the electronic spectra, d.c.~resistivity and the specific heat of disordered Kondo insulators that requires the evaluation of the conduction electron spectral function. In this section we describe the DMFT-CPA theoretical framework that we use to compute the single particle dynamics namely the non-perturbative electronic self-energy of the correlated $f$-electrons including its full frequency dependence and the electronic spectral functions in presence of Anderson-type disorder in the conduction electrons. For a clean system, the DMFT framework maps a strongly correlated lattice to an auxiliary strongly interacting impurity problem, that in this particular case, is a single impurity Anderson model (SIAM). In the SIAM we have an strongly interacting impurity, with local repulsive interactions embedded in a non-interacting host, that is determined self-consistently within a computational framework. In the presence of disorder this scheme maps the disordered lattice on to an ensemble of impurity problems each of which is embedded in a disorder averaged effective medium. In the following we outline the self-consistency equations that constitute the above scheme for temperature, $T=0$. The same self-consistent iterative scheme follows for $T>0$ as well.

The physical quantity that describes the single particle excitations in a many body system is the Green function or the `propagator'.
For a clean Kondo insulator or heavy fermion metallic system within the framework of DMFT, the $d$($f$)- electron Green's functions 
$G^d(\omega)(G^f(\omega))$ are given by,
\begin{align}
&G^d(\omega)=\left[\omega^+-\frac{V^2}{\omega^+-\tilde{\Sigma}_f(\omega)-\epsilon_f}-\Delta(\omega)-\epsilon_d\right]^{-1},
\label{Eq:Gc_app1}\\
&G^f(\omega)=\left[\omega^+-\tilde{\Sigma}_f(\omega)-\frac{V^2}{\omega^+-\Delta(\omega)-\epsilon_c}-\epsilon_f\right]^{-1},
\label{Eq:Gf_app1}
\end{align}
where, the on-site energy for the $d$- and $f$- electrons, namely, $\epsilon_d$ and $E_f$ are chosen such that the respective lattice filling 
is maintained, that in turn determines whether we obtain a Kondo insulator or a heavy fermion metal and $\tilde{\Sigma}_f(\omega)$ denotes the $f$-electron self-energy including the Hartree part. Note that we have set $\mu=0$ such that all energies are measured w.r.t. the Fermi level. 

\begin{figure}[htp!]
  \includegraphics[clip=,scale=0.6]{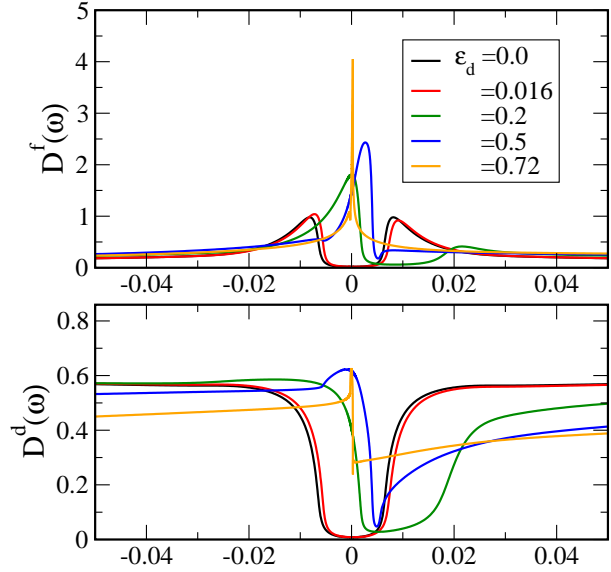}
  \caption{{\bf Density of states in presence of particle hole asymmetry} The $f$-electron (top panel) and the $d$-electron (bottom panel) density of states for various $\epsilon_d$ is plotted to demonstrate how a variation of $\epsilon_d$ may not only introduce low-lying electronic states but also induce low lying sharp features on the order of the respective Kondo scale.}
  \label{fig:DOS_ed}
\end{figure}

The quasiparticle spectra (or density of states) for a respective Kondo insulator could thus be derived to be,
\begin{align}
  &D^d(\omega)\sim \rho_0\left(\omega-\epsilon_c-\frac{ZV^2}{\omega-Z\epsilon_f^*}\right),\label{Eq:Dc}\\
  &D^f(\omega)\sim \frac{Z^2V^2}{\omega-Z\epsilon_f^*}\rho_0\left(\omega-\epsilon_d-\frac{ZV^2}{\omega-Z\epsilon_f^*}\right),
  \label{Eq:Df}
\end{align}
where, $\epsilon_f^*$ is a renormalization incurred by the bare $f$-level due to the underlying many-body dynamics. The low energy behavior embodied in Equations~\eqref{Eq:Dc} and~\eqref{Eq:Df} would be extremely important in understanding why the Kondo gap is so fragile against minute amount of disorder in $\epsilon_d$. A variation in $\epsilon_d$ not only introduces low lying electronic states  inside the gap but also generates sharp features over exponentially suppressed energy scales. As a demonstration of this, we plot $D_f(\omega)$ and $D_c(\omega)$ in Figure~\ref{fig:DOS_ed} (top and bottom panel respectively) for different values of $\epsilon_d$. 

Subsequently for a clean Kondo insulator, $G^d(\omega)=\left[\omega^+-\frac{V^2}{\omega^+-\Sigma_f(\omega)}-\Delta(\omega)\right]^{-1}$ and 
$G^f(\omega)=\left[\omega^+-\Sigma_f(\omega)-\frac{V^2}{\omega^+-\Delta(\omega)}\right]^{-1}$, where $\Sigma_f(\omega)$ is the conventional Hartree corrected conventional self-energy of the correlated $f$ electrons, being purely local or momentum independent within DMFT. The $d$-electrons being itinerant hybridize with the non-interacting (DMFT) host via the {\it true} hybridization function, $\Delta(\omega)$, while the quantity, $\frac{V^2}{\omega^+-\Delta(\omega)}=\Delta_f(\omega)$, may be thought of as an effective hybridization function for the otherwise localized $f$-electrons. Subsequently, the quantity, $\Sigma_d(\omega)=\frac{V^2}{\omega^+-\Delta(\omega)}$ may be thought of as an effective $d$-electron self energy.

In the presence of disorder, the above Greens functions are replaced by their disorder averages as the following:
\begin{align}
G^d_{ave}(\omega)&=\langle G^c_j(\omega)\rangle\nnu\\
&=\left\langle\left[\omega^+-\frac{V^2}{\omega^+-\Sigma_{fj}(\omega)}-\Delta_{ave}(\omega)-\epsilon_{dj}\right]^{-1}\right\rangle,\label{Eq:Gc_app2}\\
G^f_{ave}(\omega)&=\langle G^f_j(\omega)\rangle\nnu\\
&=\left\langle\left[\omega^+-\Sigma_{fj}(\omega)-\frac{V^2}{\omega^+-\Delta_{ave}(\omega)-\epsilon_{dj}}\right]^{-1}\right\rangle,\label{Eq:Gf_app2}
\end{align}
where $j$ represents the disorder realization and correspondingly, $\epsilon_{dj}$ represents the random $c$-electron energy drawn from some disorder distribution, $P(\epsilon_{dj})$, that in the current case is represented by a Gaussian function with a mean at $\langle\epsilon_{dj}\rangle=\epsilon^d_{ave}=0$, and the variance given by the disorder strength, $W$. Notice that in the above equations, the CPA is constituted in an arithmetic sum of the $G_{j}^d(\omega)$'s, embodied within the definition of $\left\langle\dots\right\rangle$, and also in approximating $\Delta_j(\omega)=\Delta_{ave}(\omega)$ to be the same for all the disorder realizations. 
The disorder averaged $c$- and $f$- electron spectral functions may then be evaluated as, 
$D^d_{ave}(\omega)=-\frac{1}{\pi}\mathrm{Im}G^d_{ave}(\omega)$ and $D^f_{ave}(\omega)=-\frac{1}{\pi}\mathrm{Im}G^f_{ave}(\omega)$, respectively.

The underlying lattice information is built in the equation:
\begin{align}
  G^d_{latt}(\omega)=\int_{-\infty}^\infty \frac{\rho_0 (e) de}{\omega^+-\epsilon^d_{ave}-\Sigma^d_{ave}(\omega)-e}
  =H[\gamma_{ave}],
  \label{Eq:latticeG}
\end{align}
where, a disorder averaged $c$-electron self-energy, $\Sigma^d_{ave}(\omega)$ can be extracted from Equation~\eqref{Eq:Gc_app2},and $H[\gamma_{ave}]$ is the Hilbert transform of $\gamma_{ave}=\omega^+-\epsilon^d_{ave}-\Sigma^d_{ave}$. Equation~\eqref{Eq:latticeG} helps us construct a new hybridization function, 
\begin{align}
  \Delta_{ave}(\omega)=\gamma_{ave}-1/H[\gamma_{ave}]
  \label{Eq:hyb}
\end{align}
In the present calculations we use a semi-circular density of states, corresponding to a Bethe lattice, that is represented by $\rho_0(e)=\frac{2}{\pi t}[1-(e/t)^2]^{1/2}$. This reduces, 
$\Delta_{ave}(\omega)=\frac{t^2}{4}G^d_{latt}(\omega)$. Equations~\eqref{Eq:Gc_app2}, \eqref{Eq:latticeG}, and \eqref{Eq:hyb} constitute the DMFT self-consistency equations and determine the disorder averaged effective medium. At DMFT convergence, $G^d_{latt}(\omega)=G^d_{ave}(\omega)$ within some tolerance. Nonetheless, it is worth mentioning that 
the qualitative features of results present is independent of the specific choice of the lattice.

The above flowchart hides the truly difficult part of this entire calculation: obtaining the $\Sigma_f(\omega;T)$. So, we now comment on the calculation of the interacting self-energy, $\Sigma_f(\omega)$. We use the local moment approach to compute the $f$-electron self-energy. The LMA is a diagrammatic perturbation theory based approach, built around the two broken-symmetry, local moment solutions $(\mu=\pm|\mu_0|)$ of an unrestricted Hartree-Fock mean field approximation (note that the notation for the local moment should not be confused with the chemical potential). Subsequently, the physics due to Kondo effect embodied in spin-flip dynamics are built in through an infinite-order resummation of a specific class of diagrams that quantify the transverse spin-flip processes. A low energy spin flip scale, $\omega_m$ is thereby generated and is identified through the position of the peak of the imaginary part of the transverse spin polarization propagator within this approach. Physically and quantitatively, this low energy scale is of the same order as the Kondo scale, $\omega_K$ as mentioned in this work. For more details about the local moment approach as used in impurity or clean and disordered lattice systems, we urge the readers to refer to several previous works carried out with this approach as outlined in References~\onlinecite{Eastwood1998,logan2002finitetempdynamics,vidhyadhiraja2003dynamics,Raja2_2005}
For $T>0$ the main difference occurs in the many-body diagrammatic calculation of $\Sigma_f(\omega;T\ne 0)$ for a particular disorder realisation. This involves the calculation of the finite temperature spin polarisation propagator. Full details of the structure and implementation of the generic finite $T$ and asymmetric Anderson impurity model is discussed in some of the early works~\cite{logan2002finitetempdynamics,vidhyadhiraja2003dynamics,vidhyadhiraja2004dynamics,logan2005dynamics} which the reader is referred to for further information.

For the disordered case, the calculations become enormously complex as one now needs to obtain several such Anderson impurity model solutions both at and away from particle-hole symmetry including the capturing of exponentially small Kondo scales spanning over approximately 6 orders of magnitude. It should also be noted that the conventional numerical implementation of the local moment approach discussed in the above references needs to be modified. The respective modification was developed only recently and outlined in great detail in Ref~\cite{my_paper_TMT}. We also state in passing that the local moment approach, in its current implementation, has its own drawbacks in regards to its applicability to other Kondo correlated models like the Kondo model. Nevertheless, in comparison to other state-of-the-art methods like the numerical renormalization group or the continuous time quantum Monte Carlo method, the local moment approach is best suited for the current scheme of disordered correlated models within the premises of DMFT like scenarios.

\section{Transport: d.c. conductivity}
\label{appb}
Within the framework of the DMFT, a knowledge of the one particle excitations represented 
by the Greens functions and their $(\omega,\; T)$ dependencies are sufficient to determine the transport properties. In particular, the absence of any momentum dependence in the electronic self-energy leads to the 
strict absence of any vertex corrections in the current-current correlation function. Thus within DMFT~\cite{DMFT}, the 
conductivity, for a Bethe lattice is given by, 
\begin{align}
\sigma(\omega;T)=&\sigma_0 \frac{t^2}{\omega}\int_{-\infty}^{\infty}d\omega_1 \frac{f(\omega_1)-f(\omega+\omega_1)}{\omega}\nnu\\
&\langle D^d_{ave}(\epsilon,\omega;T)\rangle_\epsilon\langle D^d_{ave}(\epsilon,\omega+\omega_1;T)\rangle_\epsilon
\end{align}
where, $\langle D^d_{ave}(\epsilon,\omega;T)\rangle_\epsilon=\int_{-\infty}^{\infty}d\epsilon\rho_0(\epsilon)D^d_{ave}(\epsilon,\omega_1;T)$, with $\sigma_0\sim 10^4-10^5\Omega^{-1}cm^{-1}$, $D^d_{ave}(\epsilon,\omega;T)=-\frac{1}{\pi}\mathrm{Im}G^d_{ave}(\epsilon,\omega;T)$ and $f(\omega)$ is the Fermi distribution function.
The d.c. resistivity can then be evaluated as, 
$\rho(\omega=0;T)\propto\left[\int_{-\infty}^{\infty}\frac{-\partial f(\omega)}{\partial\omega}(\langle D^c_{ave}(\epsilon,\omega;T)\rangle_\epsilon)^2 d\omega\right]^{-1}$. 
For the hypercubic lattice, 
\begin{align}
\sigma(\omega;T)\propto&\frac{1}{\omega}\int_{-\infty}^{\infty}d\omega_1 \frac{f(\omega_1)-f(\omega+\omega_1)}{\omega}\nnu\\
&\langle D^d_{ave}(\epsilon,\omega;T) D^d_{ave}(\epsilon,\omega+\omega_1;T)\rangle_\epsilon
\end{align}

\section{Specific heat}
\label{appc}
In the following we outline the method and assumptions underlying the calculation of specific heat for the 
system studied. The reader is also requested to see Ref.~\onlinecite{Costi_sp_ht}. Our basic assumption underlying the treatment of the disordered system, lies in mapping the 
interacting disordered system onto an ensemble of independent, Anderson impurities, each of which is embedded 
into a bath via a hybridization function, that in turn depends on $\Delta_{ave}(\omega)$ and the 
local potentials, $\epsilon_{dj}$'s, as outlined in Appendix~\ref{app1}, by equations~\ref{Eq:Gc_app2},~\ref{Eq:Gf_app2},~\ref{Eq:latticeG} and~\ref{Eq:hyb}. The specific heat for the original system is then 
calculated by taking the average over the single-impurity results with the appropriate distribution
function, which in the current case is a Gaussian. The calculation of specific heat would thus involve the 
computation of the total energies of the ensemble of Anderson impurities, such that 
$C_v(T)=\frac{\partial E_{ave}}{\partial T}$, where $E_{ave}$ is the disorder averaged energy 
of the system. In the following, we outline the equations determining the total energy calculation of 
a single impurity Anderson model.

The single impurity Anderson model is represented as,
\begin{align}
  \mathcal{H}_{SIAM}&=\sum_{k\sigma}\epsilon_{ck}c_k^\dagger c_k+\epsilon_f\sum_\sigma f_\sigma^\dagger f_\sigma+Un_{f\uparrow}n_{f\downarrow}+\nnu\\
                     &\sum_{k\sigma}V_kc_{k\sigma}^\dagger f_\sigma+h.c.,
\end{align}
where, $\epsilon_{ck}$ represents the energy dispersion of the underlying host of $c$-electrons in which a 
correlated impurity site represented by $f$ electrons is embedded via the hybridization energy denoted by 
$V_k$, such that the hybridization function is given by, $\sum_k \frac{|V|^2}{\omega^+-\epsilon_{ck}}=\Delta(\omega)$; $n_{f\sigma}$ denoted the occupancy of the $f$-electron with spin $\sigma$ on the impurity site and $U$ is the Coulomb energy cost to be paid when 
two electrons of opposite spin sit on the impurity site; $\epsilon_f$ is the on site potential energy 
of the impurity site.
\begin{figure}[htp!]
\includegraphics[clip=,scale=0.5]{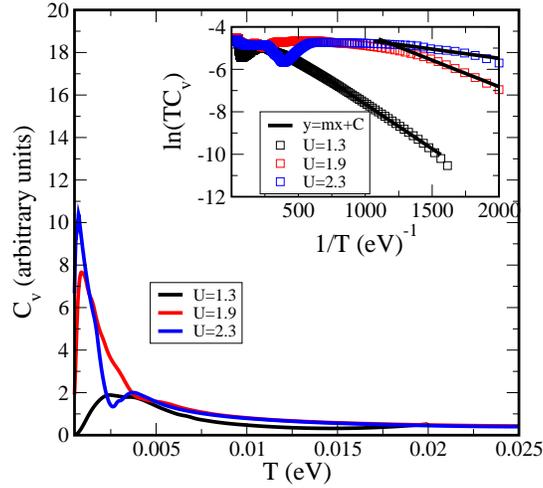}
\caption{(Main panel)The specific heat of a Kondo insulator is plotted as $C_v$ v/s $T$ on a linear scale for $U=1.3,\;1.9,\;2.3$. (Inset) $ln(C_v T)$ v/s $1/T$ is plotted to demonstrate the gradual crossover to a 
low $T$ activated behavior of $C_v$. The solid black line depicts a straight line fit. The expected 
trend, of a decreasing magnitude of the slope with an increasing $U$ is also successfully captured within the simulation.}
\label{fig:Cv_KI}
\end{figure}
\begin{align}
  E_{imp} &= \epsilon_f \sum_{\sigma} \langle n_{f\sigma}\rangle + U D +\frac{1}{2} E_{hyb} + \nnu\\
  &\frac{1}{\pi} \sum_\sigma\int d\omega n_F(\omega) Im \left\lbrack G_{f\sigma}(\omega)\omega\frac{\partial\Delta(\omega)}{\partial\omega}\right\rbrack,
\end{align}
where, $D$, is the double occupancy on the impurity site and can be represented in terms of the single
particle spectral function, $D_G^\sigma=-\mathrm{Im}G_\sigma(\omega)/\pi$ and the imaginary part of the single particle self energy,  $D_\Sigma^\sigma=-\mathrm{Im}\Sigma_\sigma(\omega)/\pi$, as the following:
\begin{align}
\langle n_{f\uparrow}n_{f\downarrow}\rangle &= \frac{\Sigma^\uparrow_{\rm Hartree}}{U} \int d\omega_1\, D^\uparrow_G(\omega_1) n_F(\omega_1) \nnu \\
&+ \frac{1}{U} \int d\omega_1\, d\omega_2 \,D^\uparrow_G(\omega_1) D^\uparrow_\Sigma(\omega_2)
 \frac{n_F(\omega_1)-n_F(\omega_2)}{\omega_1-\omega_2}\nonumber \\
&=n^\uparrow_{\rm occ}n^\downarrow_{\rm occ} + \nnu\\
&\frac{1}{U} \int d\omega_1\, d\omega_2 \,D^\uparrow_G(\omega_1) D^\uparrow_\Sigma(\omega_2)
 \frac{n_F(\omega_1)-n_F(\omega_2)}{\omega_1-\omega_2}
 \label{Eq:DO1},
\end{align}
where $n_F$ is the Fermi distribution function. The term $E_{hyb}$ represents the energy of the impurity 
due to its hybridization with the $c$-electrons and is given by,
\begin{equation}
  E_{hyb} =  -\frac{2}{\pi}\sum_{\sigma}\int d\omega\, n_F(\omega) \mathrm{Im}\lbrack\Delta(\omega)G_{d\sigma}(\omega)\rbrack
\end{equation}
As mentioned earlier, we use the local moment approach to compute the impurity self-energy and the impurity properties within the DMFT framework. While this approach comes with its limitations it's single particle properties have been extensively benchmarked with the numerically exact, numerical renormalization group calculations. For the current case of total energy calculation, and thereby the specific heat, we ensured that it reproduces the basic features in some known limits of the system under consideration. In Figure~\ref{fig:Cv_KI} we therefore plot the computed $C_v(T)$ for a clean Kondo insulator. As seen from Figure~\ref{fig:Cv_KI}, where we plot $ln(TC_v)$ vs $1/T$, the method clearly depicts the expected trends for that of a clean Kondo insulator, namely, a low $T$ activated behavior. It also correctly depicts a decreasing magnitude of the slope with increasing $U$; this slope is also a measure of the hybridization gap.

%
\end{document}